\documentstyle[prd,aps]{revtex}
\begin{document}
\wideabs{
\title{Moving Mirrors and Thermodynamic Paradoxes}

\author{Adam D. Helfer}

\address{Department of Mathematics, University of Missouri -- Columbia,
Columbia, MO 65211, U.S.A.}

\date{\today}
\maketitle
\begin{abstract}
Quantum fields responding to ``moving mirrors'' have been predicted to
give rise to thermodynamic paradoxes.  I show here that the assumption
in such work that the mirror can be treated as an external field is
invalid, and the exotic energy--transfer effects necessary to the
paradoxes are well below the scales at which the model is credible.  A
model with a first--quantized point--particle mirror is considered; for
this it appears that exotic energy--transfers are
lost in the quantum uncertainty in the mirror's state.
Examining the physics giving rise to these limitations shows that an
accurate accounting of these energies will require a model which
recognizes the mirror's finite reflectivity, and almost certainly a
model which allows for the excitation of internal mirror modes, that is,
a second--quantized model.
\end{abstract}
}
\def\T{{\widehat T}}
\def\scri{{\cal I}}
\def\dddot#1{{\partial _t^3{{#1}}}}
\def\H{{\widehat H}}
\def\e{{e}}


\section{Introduction}

Almost twenty--five years ago, Davies and
Fulling~\cite{davies76,fulling76}, following a suggestion of
DeWitt~\cite{dewitt75} (see also~\cite{moore70}), introduced the
``moving mirror'' models:  massless scalar quantum fields in
two--dimensional Minkowski space responding to perfectly reflecting
boundaries.  Such models have been of inestimable value in clarifying
conceptual issues raised by more complicated theories; most notably,
there are connections between moving mirror models and the Hawking
process.

There are still aspects of the moving--mirror models which are not
satisfactorily understood.  The most important of these are the
thermodynamic paradoxes, which seem to be consequences of basic
features of the models, and so unavoidable in them.  Consider for
simplicity the case of a non--relativistic mirror with position
$q(t)$.  Then, assuming the field was initially in the vacuum state
(and that in the far past the mirror was stationary), the expected
energy on the right (respectively, left) of the mirror is
\begin{equation}
\langle E_{{\rm right}\atop{\rm left}}\rangle
  =\mp (12\pi )^{-1}(\hbar /c)\ddot q+O(1/c^2)\, .
\end{equation}
Now we come to the key point.  It follows immediately that
the total expected energy in the
field is
\begin{equation}
\langle E_{\rm left}+E_{\rm right}\rangle =0+O(1/c^2)\, ,
\end{equation}
so (to lowest order) {\em no }
energy is required to move the mirror.
This is extraordinary.  {\em The leading effect of the
mirror's motion is to separate the vacuum into packets
of positive and negative expected energy, at no energetic
cost.  }  If these expectation--values can be regarded
as classical energies, then we have a direct violation of the 
second law of thermodynamics.  One can easily construct
paradoxes based on this, and in fact Davies 
described a perpetual--motion machine which turns around
this idea~\cite{davies82}.  (See also~\cite{ford78}.)

One might raise some objections to specific elements of Davies's
proposal, and indeed various workers have done so (mainly concentrating
on the problems of absorbing the negative energy
packets~\cite{grove86,helfer98a,helfer98b}; see also the earlier
paper~\cite{ford78}).  Still, it seems hard to avoid the central
point:  if one can split the vacuum into positive-- and
negative--energy parts, with negligible energetic input, isn't one
violating the second law? Even setting aside possible difficulties
associated with managing or absorbing the negative--energy packets,
couldn't one drive perpetual--motion machines by simply keeping the
positive energies produced by the mirror's motion, leaving the
negative--energy packets to go their ways?

The aim of the present paper is to resolve this point.  I shall show
that the approximation that has been made, that the mirror's trajectory
can be treated classically, is invalid for the purpose of computing the
necessary energy transfers.  The limitations of validity of the
classical model are reached before effects of energy--exchange between
it and the quantum field can be computed.  Thus, insofar as the mirror
can be treated classically, there is no violation of the second law.

Going beyond this classical model, I shall consider a model with a
non--relativistic, first--quantized mirror moving in an external
potential.  At least where the potential is quadratic, it will be shown
that the measurement of the energy packets is always accompanied by a
much larger spread in the mirror's energy.  This means that attempts to
measure the vacuum field energies cause quantum fluctuations in the
mirror's state, fluctuations that cannot be ignored for the purposes of
understanding the energy transfers between the field and the mirror.

It should be emphasized that the present work indicates that any
``semiclassical'' attempt to model the quantum back--reaction on the
mirror is invalid for the purposes of modeling the energy--transfers
that would occur in attempts to measure the vacuum field energy
density.  This is because semiclassical approximations (which give the
back--reaction of the quantum field on the mirror in terms of
expectations) are precisely those which assume that fluctuations in the
mirror's state are negligible, and this is just what fails here.

The present results fit well with those of a related investigation, by
Parentani~\cite{parentani96}.  He introduced a model with a
second--quantized mirror in a linear external potential.  He was able
to show (with certain approximations) that the forward quanta would
decohere.  This is because their states become correlated with that of
the mirror.  The general lesson to be drawn from the models, then, is
that the entanglement of the mirror's state with that of the field can
be a dominant effect, and the entanglement can involve fluctuations in
the mirror's energy larger than the field's energy.

One can view the present work as estimating the magnitudes of the
effects of quantum fluctuations in the mirror's state on the
energy--transfers in the system; the effects are large enough to
invalidate the external--field approximation.  However, to go beyond
this negative conclusion, and analyze in detail 
{\em what does } happen in the energy transfers, is
another issue.  I shall argue below that even this model is probably
inadequate for a satisfactory understanding of these issues, and it will be
necessary to pass to a theory where the internal degrees of freedom of
the mirror (and the scattering of virtual field quanta from these) are
accounted for.  This is surprising and perhaps
disconcerting:  one would have thought that a quantum field responding
to a slow, heavy mirror could be analyzed without needing to account for
the mirror's structure as a system of quantum fields.  But
if one wants to understand the vacuum energies, such an analysis seems
necessary.

Although such a sophisticated model will ultimately be necessary, there
are good reasons for considering the non--relativistic first--quantized
point--particle model, at least initially.  The most important one is
that there is little ambiguity in defining it, whereas to go beyond it
requires many choices.  (The more sophisticated models require one to
make assumptions amounting to a choice of dispersive susceptibility
$\chi (\omega )$, and there is functional freedom in doing so.) The
non--relativistic mirror, by contrast, can be a first--quantized point
particle, and there is little ambiguity in how to proceed.  Thus the
point--particle results, while more limited, are at least clearly
model--independent.

Another reason for starting with the first--quantized point--particle
model is that the quantum measurement issues can be analyzed at a
fairly elementary level.  Finally, the model is perhaps of some
interest beyond the present paper.  The very fact that it is of limited
validity can be turned to advantage, because higher relativistic
corrections can be ignored and a great deal of its structure can be
worked out explicitly.

In Section II, the Davies--Fulling models are reviewed.  
This section may be read rapidly, but should not be skipped.
No details of
the calculations are given, but the physical basis of the
renormalization and some of the limitations on the validity of the model
are discussed in IIB.  These limitations figure essentially in later
arguments.  Section III briefly derives the first--quantized mirror
model.  Section IV gives the main analysis of the measurement of field
vacuum energy and its limitations.  The last section summarizes the main
conclusions.

In most places, particularly in estimates of the magnitudes of energies,
factors of $c$ and $\hbar$ are given explicitly.  However, factors of
$c$ have been omitted in a few places (advanced and retarded
coordinates, etc.), where they would make the appearance of the
equations unnecessarily complicated.

\section{A Classical Moving Mirror}

In this section, I shall review the standard treatment of a massless
field influenced by a moving mirror in two--dimensional Minkowski
space~\cite{fulling76}.  No details of standard computations will be
given; the emphasis will be on the physical assumptions and
consequences.

\subsection{Basic Formalism and Results}

Let $(t,x)$ be coordinates on two--dim\-en\-sional Min\-kow\-ski space with
metric $ds^2=dt^2-dx^2$.  We introduce retarded and advanced null
coordinates by $u=t-x$ and $v=t+x$ as usual, and vectors $l^a=\partial _v$, 
$n^a=\partial _u$.  
It is convenient to regard
the trajectory of the mirror as given by $v=V(u)$ or $u=U(v)$.  We
assume that the trajectory is timelike and is asymoptotically stationary
in the past.  

We consider a massless scalar field.  Any solution to the field equation
can be written locally as $\phi (u,v)=f(u)+g(v)$.  The mirror is
considered to enforce the boundary condition $\phi (u,V(u))=0$.  Thus we
must have $f(u)=-g(V(u))$.  We shall write
\begin{equation}
 \phi =f(u)-f(U(v))\, ,\qquad v<V(u)
\end{equation}
(to the left of the mirror) and
\begin{equation}
 \phi =g(v)-g(V(u))\, , \qquad v>V(u)
\end{equation}
(to the right).  Thus the symbol $f$ will only be used for fields on the
left, and $g$ only for fields on the right.  Then 
the functions $f$ and $g$ can be considered data
at $\scri ^-$ for the field.  We may also interpret these equations at
the operator level; then $\hat f$ and $\hat g$ are the ``in'' operators.

The stress--energy operator is
\begin{equation}
 \T _{ab}=:\T _{ab}: +\langle\T ^{\rm ren}_{ab}\rangle\, ,
\end{equation}
where the colons stand for normal ordering and 
$\langle\T ^{\rm ren}_{ab}\rangle$ is the
renormalized vacuum expectation value.  This last is defined by
point--splitting.  One starts with the formal expression
\begin{equation}
  \T ^{\rm formal}_{ab}=
\left( \delta ^p_a\delta ^q_b -(1/2)g_{ab}g^{pq}\right)
\nabla _p\hat\phi \Bigr| _{(u_1,v_1)}
   \nabla _q\hat\phi \Bigr| _{(u_2,v_2)}
  \, ,
\end{equation}
and considers the limit as $(u_2,v_2)\to (u_1,v_1)$.  The
expectation value $\langle 0|\T ^{\rm formal}_{ab}|0\rangle$ contains
two terms:  a divergent one which is independent of position, and a
finite term.  It is the finite term which is 
$\langle\T ^{\rm ren}_{ab}\rangle$.
The divergent
term, present even in Minkowski space, is the ``unrenormalized
stress--energy of the Minkowski vacuum.''
The result is
\begin{equation}\label{Tren}
\langle\T ^{\rm ren}_{ab}\rangle
  =(12\pi )^{-1}\hbar\left( {3\over 4}\left( {{V''}\over{V'}}\right) ^2
  -{1\over 2} {{V'''}\over{V'}}\right) l_al_b
\end{equation}
on the right.  (On the left, one has an expression of the same form,
with $U$ replacing $V$ and $n_a$ replacing $l_a$.)  In the limit of
non--relativistic motion, with the trajectory given by $x=q(t)$, we have
\begin{equation}\label{Trena}
\langle\T ^{\rm ren}_{ab}\rangle
  =-(12\pi )^{-1}(\hbar /c^2)
  \left(  (1+\dot q)\dddot{q} +3{\ddot q}^2\dot q
\right) l_al_b+\cdots\, .
\end{equation}
(We have given as many terms as we shall need later.)
This is to be evaluated at the time $t'$ such
that $(t-t',x-q(t'))$ is null future--pointing.

From these formulas, one can derive expressions for the expected
renormalized energy in the field to the left and the right of the
mirror:
\begin{eqnarray}
  \langle E_{\rm right}\rangle&=&-(12\pi )^{-1}
     (\hbar /c^2)\int _{q(t)}^\infty
  \left( \dddot{q}  +3{\ddot q}^2\dot q
\right) \, dx\\
  &=&-(12\pi )^{-1}(\hbar /c)\int _{-\infty}^t
  \left( \dddot{q}  +3{\ddot q}^2\dot q
\right) \, dt'\\
  &=&-(12\pi )^{-1}(\hbar /c)\ddot{q}+\cdots\, .
\end{eqnarray}
(An integration by parts can be used to justify discarding the
second term when passing to the last line.)
On the left of the mirror, one has
\begin{equation}
 E_{\rm left}=+(12\pi )^{-1}(\hbar /c)\ddot{q}+\cdots \, .
\end{equation}
Thus, to lowest order, no total expected energy is produced, but the
mirror's motion effects a separation of the vacuum energy into positive
and negative terms.  The leading nontrivial contribution to the total
expected energy in the field is of order $\dot x$ (that is, $v/c$)
smaller; it is
\begin{eqnarray}
 E_{\rm total} &=&-(6\pi )^{-1}(\hbar /c^2)
     \int _{-\infty}^t \dddot{q}\dot q\, dt'\\
  &=&-(6\pi )^{-1}(\hbar /c^2)\left[ {\ddot q}{\dot q}
  +\int _{-\infty}^t ({\ddot q})^2 dt'\right]\label{Trenb}
\, .
\end{eqnarray}
Thus the total energy put into the field must be positive, if the motion
is asymptotically inertial.

\subsection{The Renormalization}

While all of the foregoing is standard, one must remember that we do
not at present have a first--principles understanding of the infinite
vacuum energy density and (therefore) of its renormalization.  While
the standard computation of $\langle\T ^{\rm ren}_{ab}\rangle$ will be
accepted here (within a regime of applicability to be discussed
shortly), since the interpretation of this quantity is critical to the
physics of the mirror, it is appropriate to discuss what has been done
carefully.
These points are important:

(a) The ``operator part'' of 
$\T _{ab}$ --- that is, the operator
modulo the addition of c--number terms like $\langle\T ^{\rm
ren}_{ab}\rangle$ --- is determined by the equation of motion for the
fields, and so is unambiguously defined irrespective of the
renormalization.  In other words, different choices of renormalization
can only affect the c--number terms.

(b) It is not trivial that the theory is renormalizable.
The idealized perfect--reflector nature of the boundary causes a great
deal of cancellation of ultraviolet contributions to the
stress--energy in the neighborhood of the
mirror.  In a more realistic model, one would expect dispersive
effects to disturb these cancellations.  This would lead to terms which
were formally divergent as one approached the mirror (although the
theory itself would break down as one approached within a distance of
the order of the skin depth of the mirror).  
Cf. ref.~\cite{helflang99}.

A related point is that we have ignored whatever internal physics the
mirror has which causes it to reflect.  For an actual (electromagnetic)
mirror, there are ions and conduction electrons whose contributions to
the electromagnetic stress--energy outside the mirror might not be
ignorable.

(c)  Consider for the moment replacing the perfectly reflecting mirror
by a more realistic model, where one has a mirror with a dispersive
susceptibility tending rapidly to zero beyond some cut--off frequency
$\omega _{\rm p}$.  The effect of this would be to introduce a
frequency--dependent potential term into the equation of motion, or
equivalently, in coordinate space, a convolution of $\hat\phi$ with the
Fourier transform of that potential.  This term would act like a
perfect reflector on field modes of frequencies $\omega\ll\omega _{\rm
p}$, but the structure of the potential would become important at
scales $\omega \sim\omega _{\rm p}$.

In such a model, the mirror will act like a classical reflector of
low--frequency modes only as long as the time scale defined by its
acceleration is significantly larger than $1/\omega _{\rm p}$.  If the
acceleration is greater, we must take into account the mixing of
low--frequency and high--frequency modes due to the mirror's motion. 

What this means for the present paper is that the computation of
$\langle\T ^{\rm ren}_{ab}\rangle$ is only credible as long as the
inverse time scales over which $\dot q$ changes are much less that the
plasma frequency $\omega _{\rm p}$ of the mirror.  In particular, we
must have $|\ddot q|\lesssim \omega _{\rm p}|\dot q|$ or we are not
justified in using the standard formula, equation~(\ref{Tren}), and its
consequences, equations~(\ref{Trena})--(\ref{Trenb}).

(d)  The usual procedure is to take the points $(u_1,v_1)$ and
$(u_2,v_2)$ separated by a small {\em imaginary } timelike interval.
This has the effect of introducing an ultraviolet cut--off.  This is
attractive, because one can then argue that the justification of the
procedure is that real experiments only probe an object up to a finite
frequency.  Also this procedure ascribes to Minkowski space--time a
(divergent) positive expected energy density, whereas real--separated
points give rise to negative energy densities.

However, this procedure requires one to consider the world--line
$V(u+i\delta u)$ at complex points as well, and it is hard to give a
physical interpretation of this.  If $V$ is analytic, of course, one
has a clear candidate definition for $V(u+i\delta u)$.  However, even
in this case $V(u+i\delta u)$ depends non--locally on the real
trajectory $V(u)$.  It is in particular hard to see how to reconcile
one's notions of causality (being able to change $V(u)$ freely in the
future of $u=u_0$, irrespective of its behavior in the past) with the
requirement of analyticity.

In practice, this point is usually ignored, and $V(u+i\delta u)$ simply
represented by a Taylor series whose convergence is not questioned.  We
remark that if $V(u)$ is not analytic but Cauchy's formula is used to
provide a candidate definition for $V(u+i\delta u)$, the c--number term
$\langle\T ^{\rm
ren}_{ab}\rangle$ becomes divergent; the theory is not renormalizable.

We shall not pursue this question of how or whether the standard
renormalization is justified.  Still, it is a point which is not really
satisfatorily understood.

\section{A First--Quantized Mirror}

In order to estimate the effects of quantum fluctuations in the state
of the mirror on the energy exchanges between it and the field, we must
quantize the mirror.  We shall consider a simple model, in which the
mirror is considered to be heavy and its motion non--relativistic.
Then the mirror may be treated as a first--quantized particle.  Let us
begin by anticipating the limitations of this model.

(a) If the mirror's mass is $m$, then the model will only be valid for
field modes of frequencies $\ll mc^2 /\hbar$.  The mass provides an
effective ultraviolet cut--off.

(b) The model can accurately predict dynamics only for a finite time. 
This is because eventually relativistic corrections to phases become
significant.  Correspondingly, there will be a limit to the accuracy
of the energy levels predicted by the model.

(c) It will be most important to recall that a first--quantized model
is only valid at length scales greater than the Compton wavelength
$\hbar /(mc)$.  At smaller length scales, attempts to measure the
position of the particle require localized energies large enough that
pair creation (here, of quanta of the ``mirror'' field) becomes
non--negligible, and this precisely means that the first--quantized
model breaks down.  This means that the position operator $q$ of the
mirror only has a well--defined correspondence with physical reality on
greater length scales.

(d) Even on the one--particle Hilbert space, relativistic corrections
make the inner product $\langle\psi |\psi\rangle$ non--local with a
length scale of order $\hbar /(mc)$.  This means that, as far as the
one--particle model makes sense, the quantum observable $q$ always
has a spread of at least the order of the Compton wavelength.

The general strategy will be to first consider the mirror as classical
and moving in a specified external potential ${\cal V}(q)$, and then
promote the mirror's position $q$ and momentum $p$ to quantum operators.
The Hamiltonian of the mirror is just $p^2 /(2m) +{\cal V}(q)$, so the
main work involved is to compute the Hamiltonian of the field.

In fact, for the purposes of the present paper, it is only really
necessary to compute the contributions to the vacuum energy part of the
Hamiltonian:  the normal--ordered terms are not needed.  Still, we shall
give these terms, for the purpose of making clear just what the model
is.  The dynamical consequences of the terms will be investigated
elsewhere.

We begin by working out the contributions to the field Hamiltonian 
at $t=0$ from
the left and the right of the mirror.  
(The choice $t=0$ is of course conventional; other choices of time slice
will be related by unitary transformations.)
We have
\begin{equation}
 \H _{{\rm right}\atop{\rm left}} =:\H _{{\rm right}\atop{\rm left}}:
  \mp (12\pi )^{-1}\ddot q\, .
\end{equation}
Using the mirror's equation of motion, we will replace $\ddot q$ by
$-(1/m){\cal V}'$.  For the normal--ordered terms, we have
\begin{equation}
 :\H _{\rm right} :=\int _{q(0)}^\infty :\left( {\hat g}'(-x)\right)^2 
      +\left( V'(-x){\hat g}'(V(-x))\right) ^2:\, dx\, ,
\end{equation}
with $:\H _{\rm left}:$ given by a similar expression.  We now re--write
the contribution from the second term, in two steps.  We have
\begin{equation}
\int _{q(0)}^\infty :\left( V'(-x){\hat g}'(V(-x))\right) ^2:\, dx
  =\int _{-\infty}^{q(0)} :{\hat g}'(v)^2: V'\, dv\, .
\end{equation}
Since $V'$ is a perturbation of unity, we split off a term where $V'$ is
replaced by unity, and combine it with the first term in $:\H _{\rm
right}:$ to give the Hamiltonian of a free field (in the presence of a
fixed mirror) plus a perturbation, which is
\begin{eqnarray}
 \int _{-\infty}^{q(0)} &&:{\hat g}'(v)^2: (V'-1)\, dv \nonumber\\
  &&=2\int _{-\infty} ^0:{\hat g}'(t+q(t))^2:\dot q (t)\, dt\, .
\end{eqnarray}
Thus we have
\begin{equation}
 :\H _{\rm right}: =:\H _{\rm right,\ fixed}:
  +:\H _{\rm right,\, pert}:\, ,
\end{equation}
where
\begin{equation}
:\H _{\rm right,\ fixed}: =\int _{-\infty} ^\infty
  :{\hat g}'(x)^2:\, dx
\end{equation}
and 
\begin{equation}
:\H _{\rm right,\, pert}: =2\int _{-\infty}^0 :{\hat g}'(t+q(t))^2:
  \, \dot q (t)\, dt\, .
\end{equation}
This term is already of order $v/c$.  Thus we may compute $q(t)$ and
$\dot q(t)=p(t)/m$ to the required accuracy from the mirror's
Hamiltonian
\begin{equation}
 \H _{\rm mirror}=p^2/(2m)+{\cal V}\, .
\end{equation}
Using this, choosing a Hermitian factor--ordering, and abusing notation
by keeping the same symbol for the Hamiltonian with quantized $p$ and
$q$, we find
\begin{eqnarray}\label{dromedary}
 &&:\H _{\rm right,\, pert}: =m^{-1}\int _{-\infty}^0
  e^{i\H _{\rm mirror} t}\nonumber\\
  &&\qquad \times\left(:{\hat g}'(t+q(0))^2:
       \, p(0) 
        +p(0): {\hat g}'(t+q(0))^2: \right) \nonumber\\
   &&\qquad \times e^{-i\H _{\rm mirror}t}\, dt\, .
\end{eqnarray}
This is the final expression for the 
normal--ordered part of the correction to the free--field
Hamiltonian in the model with first--quantized mirror.
As mentioned above, we do not really need this explicit form in what
follows, but present it for the purposes of defining the model.

Before analyzing how passage to this model affects the paradoxes of the
classical mirror, a few comments about the model's structure are in
order.  One can regard this model as a perturbation of a stationary
classical mirror, the perturbation parameter being $m^{-1}$.  Adopting
this point of view, one can ask how the eigenstates of the classical
mirror are affected by taking into account its finite mass.  The
integrals over the half--line in equation~(\ref{dromedary}) will contain
creation$\otimes$creation and annihilation$\otimes$annihilation
operators, and these will result in a ``dressing'' of the states.  In
particular, the vacuum state will be dressed with two--particle
contributions.  The mirror, too, will be affected by the operators $p$
and $q$; the dressing will contribute states which in the unperturbed
theory would be excited.

\section{Measurement of the Vacuum Energy}

We now take up the question of how well the vacuum energy on either side
of the mirror can be measured, and to what extent those measurements are
compatible with the treatment of the mirror as a classical object. 

Throughout this section, we consider the measurement of $\H _{\rm
right}$.  This means a measurement is made of the field energy
on the entire half--space to the right of the mirror.  (Of course, this
is for many purposes an idealization.  In many cases, one would consider
the energy--content over a fixed region of space, and restrict the
mirror to be on one side of that.  However, such analyses are cumbersome
and will not be attempted here.)  We also assume that the field is
initially in the vacuum state.


\subsection{The Classical Model}

In this subsection, we shall assume that the mirror is in a state which
can be well--modeled by a classical trajectory.  Thus we may assume that
at any time $t$ the mirror's position and momentum may be measured to
classical accuracies $\Delta q$ and $\Delta p$ which are larger than the
spreads in the corresponding quantum observables.
Then
there is a classical limit to the accuracy to which the mirror's energy
is known:
\begin{equation}
 \Delta H_{\rm mirror} \simeq
  {p\over m}\Delta p +{\cal V}'(q)\Delta q\, .
\end{equation}
In particular, the limit of the accuracy in the energy due to the classical
uncertainty in position must satisfy
\begin{equation}
  |\Delta H_{\rm mirror}|\gtrsim |{\cal V}'\Delta q|\, .
\end{equation}
However, note that $\Delta q$ must be far larger than the mirror's
Compton wavelength for the mirror to be in a classical state.  Thus we have
\begin{equation}
 |\Delta H_{\rm mirror}|\gg \langle \H _{\rm right}\rangle\, .
\end{equation}
In other words, to the extent that the classical model of the mirror's
trajectory is credible, the lack of accuracy in 
knowledge of the mirror's energy must be
far larger than the vacuum energy in the field.  This means that
while the mirror's motion splits the vacuum into energy packets of
opposite signs, the uncertainty in the energetic cost of this separation
is far larger than the magnitude of the separation itself.  Thus there
is no detectable violation of the second law.

Note too that this means an attempt to consider a term like $\langle\H
_{\rm right}\rangle$ as a semiclassical contribution to the mirror's
energy is misguided.  It is not of itself wrong, but it is a correction
far below the scale at which any classical treatment of the mirror is
valid.

\subsection{The First--Quantized Model}

In the previous subsection, we saw that energies of the scale $\langle\H
_{\rm right}\rangle$ were far below contributions which could be
meaningfully treated by a classical model of the mirror.  This of course
suggests that we must pass to a quantized mirror to understand the
energy--transfers between it and the field.  I shall do so here, using
the first--quantized model, but I shall not attempt a full analysis of
the problem.  This is partly because of technical difficulties in the
first--quantized model (as I shall explain), but there is a deeper
reason.  

We saw in the previous subsection that the mirror's Compton
wavelength entered in limiting the validity of the classical model. 
This length is the scale at which a first--quantized treatment breaks
down, so we may expect that even the first--quantized model will be
inadequate.  This is indeed the case, as will be discussed below. 
However, the analysis of the first--quantized mirror will uncover a new
physical effect in the energetics, and so we take it up here.

We have $\langle\H _{\rm right}\rangle =(12\pi m)^{-1}{\cal V}'(q)$. 
This means a measurement of the vacuum energy is essentially a
measurement of $q$.  (A strictly linear potential ${\cal V}=$const$\cdot
q$ is excluded for several reasons.  The most important of these is that
the corresponding classical trajectories would not obey the boundary
conditions necessary for the derivation of the formulas for $\langle \T
_{ab}^{\rm ren}\rangle$.)  A measurement of $q$ is always made with a
quantum uncertainty, and insofar as the first--quantized model is valid
the spread in the quantum observable must be larger than the Compton
wavelength:
\begin{equation}
 \Delta q \gtrsim \hbar /(mc)\, .
\end{equation}
Note that while the symbol used ($\Delta q$) is the same as in the
previous subsection, the meaning here is different.  Here $\Delta q$
represents not just a lack of knowledge or of measurement resolution, but
the spread of the components of the wave function with respect to the
spectral resolution of the operator $q$.

The spread in the mirror's potential energy is 
\begin{equation}
 \Delta {\cal V}\simeq {\cal V}'\Delta q\gtrsim {\cal V}' (\hbar
/(mc))\, ,
\end{equation}
which is far larger than the vacuum energy.  This suggests the relation
\begin{equation}\label{relat}
  \Delta E_{\rm mirror}\gtrsim\langle\H _{\rm right}\rangle \, ,
\end{equation}
that is, the spread in the mirror's energy must be larger than the
vacuum energy in the field.  This would mean that the vacuum energy
could not be usefully separated from the mirror's own energy, and indeed
the vacuum energy would have to be considered as part of the
constitutive energy of the mirror.

Of course, the relation~(\ref{relat}) has not been established
rigorously, because we have neglected possible cancellations between
the spreads of the mirror's kinetic energy and its potential energy.  A
careful argument would seem to be technically very difficult,
especially as we have made essentially no restriction on the
potential.  However, the physical conclusion --- that the spread in the
mirror's energy is large compared to the vacuum energy --- seems
suggestive eneough that it is worthwhile to raise as a generic
possibility.

We {\em can} establish the relation~(\ref{relat}) 
in the case of a mirror moving in a
quadratic potential ${\cal V}=kq^2/2$.  In
this case, the vacuum energy is $(12\pi m)^{-1}kq$, so a measurement of
this is precisely a measurement of $q$.  A reliable measurement of this
energy therefore requires a measurement with nominal value $\overline q$
and spread $\Delta q$ related by $\Delta q /\overline q <1$.  Now, since
$[\H _{\rm mirror},q]=i\hbar p/m$ we must have 
$\Delta E\Delta q\gtrsim \hbar\overline p /m$.  On th other hand, we
have $\Delta p\gtrsim\hbar /\Delta q$ and thus $\overline p \gtrsim
\hbar /\overline q$.  Thus the mirror's energy spread 
\begin{eqnarray}
 \Delta E&&\gtrsim \hbar /(m\overline q\Delta q)\\
  &&\gtrsim \hbar /(m {\overline q}^2)\\
  &&\gtrsim (\hbar /m) (mc/\hbar ) (1/\overline q )\\
  &&\gtrsim (\hbar /m) (k{\overline q}^2/\hbar c) (1/\overline q )\\
  &&=(\hbar /mc) k\overline q\, ,
\end{eqnarray}
which is the vacuum energy.

\subsection{Limitation of the First--Quantized Model}

The treatment of the mirror as a first--quantized particle is only
accurate within certain regimes.  An important limitation is that a real
mirror does not reflect all frequencies perfectly, but becomes
transparent to sufficiently high modes.  

To understand this, let $\omega _{\rm p}$ be a ``plasma'' frequency,
giving a scale beyond which the mirror becomes essentially transparent. 
Associated with this is a ``skin depth'' $c/\omega _{\rm p}$ to which
modes penetrate before being reflected.  The mirror's position, as a
reflecting surface, is not defined more accurately than this skin depth.
This means that the model is only credible insofar as it depends on
spatial resolutions $\gtrsim c/\omega _{\rm p}$.  Now for a realistic
mirror we must have
\begin{equation}
  \hbar\omega _{\rm p}\ll mc^2\, ,
\end{equation}
that is, the plasma energy should be less than the rest energy, or
equivalently, the skin depth should be much larger than the Compton
wavelength.  However, the vacuum energy on one side of the field is
\begin{equation}
  \left| (12\pi )^{-1} {\cal V}' {\hbar\over{mc}}\right|
  \approx (12\pi )^{-1}\left| {\cal V}\left( q+{\hbar\over{mc}}\right) 
  -{\cal V}(q)\right|
\end{equation}
is a measurement of the difference of potential energies over the
Compton wavelength $\hbar /(mc)\ll c/\omega _{\rm p}$.  Thus this energy
difference is well below the ambiguity in the mirror's Hamiltonian
\begin{equation}
 \H _{\rm mirror} ={{p^2}\over{2m}} +{\cal V}\, .
\end{equation}

We see that the first--quantized model is not accurate enough to
determine whether there are exchanges of energy between the mirror and
the field of the same scale as the vacuum energy.  The limitation is in
the treatment of the mirror as perfectly reflecting, and the neglect of
whatever internal physics of the mirror gives rise to that refelection. 
Presumably, an accurate model will require passing beyond this.

\section{Summary and Implications}

\subsection{Summary} 

I have re--examined the ``moving mirror'' models of
Davies and Fulling, giving attention to their limits of validity in
computing energy transfers between the mirror and the vacuum energy of
the field.  Insofar as the mirror can be modeled by a classical point
particle, we find that the lack of accuracy in its energy far exceeds
the vacuum energy.  This means that, while the motion of the mirror
splits the vacuum into positive and negative energy packets, the
magnitudes of those energies are far below the uncertainty in the
mirror's energy.  Thus no violation of the second law arises.

Moving beyond the classical model to first--quantized mirror, we were
able to show, at least in the case of a quadratic external potential,
that the quantum spread in the mirror's energy must be greater than the
field's vacuum energy.  This would mean that a measurement of the field
energy would necessarily drive the mirror into a superposition of
energy states, with width greater than the vacuum energy.  For more
general potentials, we gave suggestive but not conclusive arguments for
the same conclusion.

These conclusions are consonant with the results of
Parentani~\cite{parentani96}.  With a different although related
purpose, he investigated a second--quantized mirror model with certain
approximations, and found correlations between the mirror state and
those of forward--scattered quanta.  Taken together, the two models
show:  (a) that when vacuum field energies are measured, the
entanglement of the mirror's state with that of the field may be a
dominant effect; and (b) the entanglement may involve fluctuations in
the mirror energy grater than the field's vacuum energy.

Even the model of the mirror by first--quantized point particle turned
out to be insufficiently accurate for quantitative analysis of the
energy transfers between the mirror and the field.  
We found that in order to reliably compute mirror energies to a
resolution of the order of the vacuum energy, one will need to take into
account the finite reflectivity of the mirror, and its structure on
scales of the order of its skin depth.

We found too evidence for another limitation on energy measurements,
deeper than that set by the finite reflectivity.
At every point where we used the skin depth to restrict the limits of
measurability of energies, we also used the Compton wavelength
$\hbar /(mc)\ll c/\omega _{\rm p}$ of
the mirror.  The Compton wavelength is the scale at which the
first--quantized, non--relativistic model of a particle (here, the
mirror) breaks down (irrespective of its reflective properties).  The
appearance of this scale seems to indicate that in a deep way we must
confront the infinitely many degrees of freedom of the mirror as a
second--quantized object, before we will be able to have a satisfactory
understanding of the transfers of energy between it and the field.

These requirements to pass to a very deep model of the mirror in order
to reliably study the energetics of the system must be considered a
surprise, because for a long time it has been assumed that at least for
sufficiently heavy mirrors a classical model should be valid.  However,
we see that not only does this fail, but even a first--quantized mirror
model is not sufficiently refined to make positive predictions.

\subsection{Implications}

These results have serious implications for attempts to understand the
back--reaction of the quantum field on the mirror.  Even the first
conclusion, that such back--reaction effects are below the range of
applicability of the classical model, is important.  It shows that any
attempt to treat these back--reaction effects semiclassically is
misguided, because the back--reactions are far smaller than the scales
at which the classical model can be trusted anyway.

The second result (that the spread in the mirror's energy must be
greater than the vacuum energy) shows that the semiclassical model is
not merely insufficiently refined:  its basic assumption is wholly
misdirected.  A semiclassical approximation precisely assumes that the
quantum fluctuations in the mirror's state are negligible:  but the
opposite is the case.

It is hoped that the features uncovered in this simple model will
be guides to the analysis of more complicated and realistic field
theories.  In particular, a main motivation for this work was as a
warm--up for an analysis of similar issues in the Hawking process, where
a wholly convincing understanding of the backreaction has yet to be
reached.

\end{document}